\documentclass[aps,prl,twocolumn,amsmath,amssymb,amsfonts,floatfix,nofootinbib,notitlepage,showpacs]{revtex4-2}
\usepackage{times,graphics,graphicx,bm,bbm,color,xcolor,eurosym,xurl,dsfont,mathtools,physics}
\usepackage{hyperref}
\hypersetup{breaklinks=true}
\hypersetup{colorlinks,linkcolor={blue},citecolor={blue},urlcolor= {blue}}
\usepackage[T1]{fontenc} 
\usepackage{newtxmath} 
\usepackage{newtxtext} 
\DeclareFontFamily{OT1}{pzc}{}
\DeclareFontShape{OT1}{pzc}{m}{it}
              {<-> s * [1.25] pzcmi7t}{}
\DeclareMathAlphabet{\mathpzc}{OT1}{pzc}
                                 {m}{it}

\usepackage[normalem]{ulem} 
\usepackage{cancel} 
\usepackage{soul} 
\setstcolor{red}
\DeclareUnicodeCharacter{200E}{}
\newcommand{\ignore}[1]{}
\begin{document}

\title{Exact High-Temperature Quantum Area Law}

\author{A. Yousefi}
\email{ahmad.yousefi@staff.sharif.edu}
\affiliation{Department of Physics, Sharif University of Technology, Tehran 14588, Iran}

\author{A. T. Rezakhani}
\email{rezakhani@sharif.edu}
\affiliation{Department of Physics, Sharif University of Technology, Tehran 14588, Iran}
\date{\today}

\begin{abstract}
We prove a $\beta^{2}$ area law for the quantum mutual information of thermal states of local lattice Hamiltonians: for any bipartitioning $A|B$ and all inverse temperatures $\beta$ below a critical value $\beta^{*}$, we show $\mathpzc{I}_{\beta}(A,B) \leqslant \tilde{f}(\beta)\, \beta^{2}\,|\partial_{AB}|$, where $|\partial_{AB}|$ denotes the size of the boundary region and $\tilde{f}(\beta)$ remains finite as $\beta \to 0$. This quadratic scaling improves the relationship between high-temperature correlations and macroscopic thermodynamics. While a linear area law implicitly  permits a loose, temperature-independent upper bound on extractable work and binding energy, the quadratic scaling ensures that the nonequilibrium free energy available after decoupling regions and the total energy of the interface both are proportional to the area of the interface of $A|B$ and vanish as $\beta$ at high temperatures. These results show that the high-temperature mutual information and correlation are suppressed, respectively, quadratically and linearly in inverse temperature, which are consistent with the expected thermal decoupling in macroscopic thermodynamic quantities. 
\end{abstract}
\maketitle

\textit{Introduction.---}The equilibrium state of a quantum system with local quantum Hamiltonian $H$ on a $D$-dimensional lattice at inverse temperature $\beta$ ($\geqslant 0$) is the Gibbs state $\textbf{\textit{g}}_{(\beta,H)} = e^{-\beta H} / \mathrm{Tr}[e^{-\beta H}]$. A well-known result in quantum statistical mechanics is the \textit{thermal area law} \cite{wolf2008area, unreliability}, stating that for any bipartitioning of the lattice into regions $A$ and $B$, the quantum mutual information is bounded by
\begin{equation}
\mathpzc{I}_{\beta}(A,B)\leqslant 2\beta \,\Vert H_{\partial}\Vert \propto\beta \,|\partial_{AB}|,
\label{lin-AL}
\end{equation}
where $H_{\partial}$ is the $AB$-boundary interaction, $|\partial_{AB}|$ denotes the size of the boundary region, and $\Vert \cdot\Vert$ is the standard operator norm. This linear scaling in $\beta$ appears in entanglement entropy \cite{Eisert2010}. Although at low temperatures refinements of the area law  have been reported \cite{alhambra2023quantum, kuwahara2021}, at high temperatures the linear scaling has been noted to be optimal \cite{kuwahara2021}. But this linear bound is loose at high temperatures because it sets $\Vert \mathrm{Tr}_{B}[\textbf{\textit{g}}_{(\beta,H)}]\otimes \mathrm{Tr}_{A}[\textbf{\textit{g}}_{(\beta,H)}] -\textbf{\textit{g}}_{(\beta,H)}\Vert_{1}\leqslant 2$ for the trace norm ($\Vert X\Vert_{1}=\mathrm{Tr}[\sqrt{X^{\dag}X}]$), disregarding its temperature dependence and its decline at the high-temperature limit.

A similar issue also afflicts the bound on extractable work from decoupling a bipartite lattice. If one dynamically quenches the system by decoupling the spatial regions, the system is left in an out-of-equilibrium state. The maximum extractable work from this state is bounded by the free energy difference,
which gives $W_{\max}\leqslant \mathrm{Tr}[H_{\partial}(\textbf{\textit{g}}_{(\beta,H_{A})} \otimes \textbf{\textit{g}}_{(\beta,H_{B})}-\textbf{\textit{g}}_{(\beta,H_{AB})})]$. Here it is evident that the naive bound $\Vert \textbf{\textit{g}}_{(\beta,H_{A})} \otimes \textbf{\textit{g}}_{(\beta,H_{B})}-\textbf{\textit{g}}_{(\beta,H_{AB})}  \Vert_{1}\leqslant 2$ returns a nonvanishing upper bound on the extractable work at the high-temperature limit. This allows for the possibility that an arbitrarily hot system could store a finite, surface‑sized amount of extractable work. However, from a physical perspective, thermal fluctuations disrupt correlations as $\beta \to 0$, so the extractable work is expected to vanish. Thus, the linear upper bound is too loose to capture the expected decay and results in a discrepancy.

A perturbative expansion of $\mathpzc{I}_{\beta}$ around the infinite temperature suggests that the leading order is $O(\beta^{2})$. However, such an expansion is complicated by volume $V$‑dependent coefficients scaling as $O(\beta^{3} V)$ and beyond, making it challenging to extract a general, volume-independent (area) law. In this paper, we prove an exact nonperturbative, quadratic area law, valid for all local lattice Hamiltonians and for inverse temperatures below a critical value $\beta^{*}$ which only depends on the geometry and strength of the interactions. Specifically, we prove
\begin{equation}
\label{arealaw-}
\mathpzc{I}_{\beta}(A,B)\leqslant \zeta h^{2}\,f(\beta)\,\beta^{2}\,|\partial_{AB}|,
\end{equation}
where $\zeta $ is a geometric constant defined later, $h$ is the maximum interaction strength, and $f(\beta)$ is an analytic prefactor that remains finite as $\beta \to 0$ and diverges at $\beta^{*}$. We also show that this bound yields a corresponding limit on the thermodynamic cost of decoupling two partitions $A|B$ of a bipartite system, 
\begin{equation}
\label{wmaxb}
W_{\max} \leqslant \zeta h^{2} \,f(\beta)\,\beta \, |\partial_{AB}|,
\end{equation}
which scales as $\beta$ at high temperatures and resolves the thermodynamic discrepancy of earlier bounds. Further, using the same toolkit, we show that the binding energy of any bipartitioning $A|B$ obeys a similar area law as $W_{\max}$,
\begin{equation}
\label{binden}
\Delta U_{AB}\leqslant 2 \zeta h^{2} \,f(\beta)\,\beta \, |\partial_{AB}|.
\end{equation}

Traditionally, the thermal area law is regarded as a boundary-vs-bulk matter. In this framework, one asks how the boundary interaction correlates with the bulk, and one would expect the temperature dependence of the mutual information to come from boundary-bulk covariances. Our analysis shows that this expectation is not the one that leads to the improved mutual information area law. What the boundary-bulk structure actually gives is a subsystem thermality surface law---the deviation of a region's reduced state from its local Gibbs state---which is a distinct statement, linear in $\beta$, and not the mutual information area law. Our crucial observation is that the hidden $\beta$ dependence is extractable by shifting to a boundary-boundary perspective. This progress becomes possible by expressing the free energy difference bound on the mutual information as the generalized covariance of the boundary interaction with itself. We then employ the ``high-temperature exponential clustering theorem'' \cite{kliesch2014locality}, which allows to bound boundary-boundary correlations (covariances between local terms of $H_{\partial}$) diminish exponentially with distance along the interface. We geometrically sum this exponential clustering across concentric spatial shells to avoid the volume-dependent terms present in standard series expansions. 

The remainder of the paper is organized into two main steps. First, we establish an exact, nonperturbative bound on the lattice partitioning mutual information, where a straightforward exact covariance identity is discussed and we prove the quadratic area law for mutual information by grouping the boundary terms into concentric shells and using bilinearity of the covariance together with the clustering theorem. The $\beta^{2}$ scaling emerges from the covariance identity (introduced later) applied to the boundary interaction itself; the shell grouping is what reduces the double sum over the boundary terms from an area-squared to an area law. The crucial observation is that exponential decay of the covariance suppresses the volume growth of the shells: as one moves outward from a fixed boundary term, the number of terms in a shell grows polynomially while their covariance with the fixed term decays exponentially, so the sum converges to a finite value independent of the boundary area. 

This is analogous to the short-range-vs-long-range distinction in statistical mechanics, where a potential steeper than the Coulomb's is required for a thermodynamically acceptable bulk energy. Here it is the exponential clustering that keeps the boundary–boundary sum finite; otherwise, the bound would scale as $|\partial_{AB}|^{\alpha}$ with $\alpha>1$. As an immediate consequence we also derive a linear (in $\beta$) bound on the marginal and thermal factorizations of bipartite lattices using the Pinsker inequality. In addition, we apply a similar mathematical machinery to show that the decoupling work and the binding energy obey a temperature-dependent area law, consistent with thermodynamic expectations.
\\

\textit{Setting.---}Consider a bipartite quantum system on a $D$-dimensional lattice $\Lambda$ with finite-dimensional local Hilbert spaces (e.g., spins), equipped with the usual square graph distance and partitioned into regions $A$ and $B$, governed by a local Hamiltonian $H_{AB} = \textstyle{\sum_{\mathbbmss{Z} \subseteq \Lambda}} h_{\mathbbmss{Z}}$, where $h_{\mathbbmss{Z}}|_{\mathrm{diam}(\mathbbmss{Z})> R}=0$. The geometric boundary corresponds to the set of sites defined as $\partial_{AB}=\lbrace i\in A\,|\,\mathrm{dist}(i,B)=1 \rbrace$, whose cardinality is $|\partial_{AB}|$. The interaction Hamiltonian 
\begin{equation}
H_{\partial} = \textstyle \sum_{\mathbbmss{Z} \text{ crosses }\partial_{AB}} h_{\mathbbmss{Z}}
\end{equation}
consists of local terms crossing this boundary, i.e., $\mathbbmss{Z}$'s with $\mathbbmss{Z} \cap A \neq \varnothing \land \mathbbmss{Z} \cap B \neq \varnothing$. We also define the interaction constant $\zeta =\max_{i\in \Lambda}|\lbrace  \mathbbmss{Z} \ni i\, |\, h_{\mathbbmss{Z}}\neq 0 \rbrace|$, which bounds the maximum number of interaction terms that can involve any single site. We assume $\lVert h_{\mathbbmss{Z}} \rVert \leqslant h$, for all $\mathbbmss{Z}$. 

Let the equilibrium (Gibbs) state of the composite system be $\boldsymbol{\varrho}_{AB}=\textbf{\textit{g}}_{(\beta,H_{AB})}$, from which the reduced states are given by $\boldsymbol{\varrho}_{A}= \mathrm{Tr}_{B}[\textbf{\textit{g}}_{(\beta,H_{AB})}]$ and $\boldsymbol{\varrho}_{B} = \mathrm{Tr}_{A}[\textbf{\textit{g}}_{(\beta,H_{AB})}]$. In addition, consider the Gibbs states of the decoupled regions $\textbf{\textit{g}}_{(\beta,H_{A})}$ and $\textbf{\textit{g}}_{(\beta,H_{B})}$, where $\textbf{\textit{g}}_{(\beta,H^{0}_{AB})}=\textbf{\textit{g}}_{(\beta,H_{A})} \otimes \textbf{\textit{g}}_{(\beta,H_{B})}$ and $H^{0}_{AB}=H_{A} + H_{B}$ is the Hamiltonian of the decoupled system.
\\

\textit{Covariance and subsystem thermality.---}To evaluate the impact of the boundary interaction $H_{\partial}$, we use a straightforward identity \cite{kliesch2014locality}. Let $H^{s}_{AB}=H_{AB}^{0}+sH_{\partial}$ (for $s\in[0,1]$) be an interpolating Hamiltonian and $\textbf{\textit{g}}_{(\beta,H_{AB}^{s})}$ be its corresponding Gibbs state. By using the fundamental theorem of calculus the difference in the expectation values of any observable $O$ between the coupled and decoupled states becomes
\begin{equation}
\label{eq:pertformula}
\begin{split}
&\mathrm{Tr}[O(\textbf{\textit{g}}_{(\beta,H^{0}_{AB})}-\textbf{\textit{g}}_{(\beta,H_{AB})})]= \beta\,\textstyle{\int_{0}^{1} ds \int_{0}^{1} dr}\\
&\,\,\times \mathrm{cov}^{r}_{\textbf{\textit{g}}_{(\beta,H_{AB}^{s})}}(H_{\partial},O),
\end{split}
\end{equation}
where $\mathrm{cov}^{r}_{\boldsymbol{\varrho}}(X,Y) = \mathrm{Tr}[\boldsymbol{\varrho}^{r} X \boldsymbol{\varrho}^{1-r}Y] - \mathrm{Tr}[\boldsymbol{\varrho} X]\, \mathrm{Tr}[\boldsymbol{\varrho} Y]$ is a (generalized) covariance. An immediate consequence is the subsystem thermality area law bounding the deviation of the local reduced state $\boldsymbol{\varrho}_{A}$ from the local Gibbs state $\textbf{\textit{g}}_{(\beta,H_{A})}$ as 
\begin{align}
\Vert \boldsymbol{\varrho}_{A}-\textbf{\textit{g}}_{(\beta,H_{A})}\Vert_{1}=&\, \textstyle{\sup_{\Vert O_{A}\Vert \leqslant 1}} | \mathrm{Tr}[O_{A}(\textbf{\textit{g}}_{(\beta,H_{AB})}-\textbf{\textit{g}}_{(\beta,H^{0}_{AB})})]| \nonumber\\
\leqslant &\, \beta\, \textstyle{\sup_{s,r,\Vert O_{A}\Vert \leqslant 1}} |\mathrm{cov}^{r}_{\textbf{\textit{g}}_{(\beta,H_{AB}^{s})}} (H_{\partial},O_{A})| \nonumber\\
 \leqslant &\, \zeta h\, \beta\, |\partial_{AB}|, \label{rhog}
\end{align} 
where we used $\|H_\partial\|\leqslant \zeta h\, |\partial_{AB}|$ and bounded the covariance trivially by the operator norms $|\mathrm{cov}_{\boldsymbol{\varrho}}^{r}(X,Y)|\leqslant \Vert X\Vert \Vert Y\Vert $. Note that this subsystem thermality area law is a purely kinematic consequence of treating the interaction as a boundary-bulk perturbation. To get past this kinematically driven area law and capture the thermodynamic scaling, we need to transition to the boundary-boundary perspective and account for the spatial structure of the covariance.
\\

\textit{Thermodynamic bound on mutual information.---}It is straightforward to bound the mutual information, 
\begin{equation*}
\begin{split}
\mathpzc{I}_{\beta}(A,B)=&\, S(\boldsymbol{\varrho}_{AB}\Vert\boldsymbol{\varrho}_{A}\otimes\boldsymbol{\varrho}_{B}) \leqslant S(\textbf{\textit{g}}_{(\beta,H_{AB})}\Vert \textbf{\textit{g}}_{(\beta,H^{0}_{AB})})\\
\leqslant&\, \beta \, \mathrm{Tr}[H_{\partial}\, (\textbf{\textit{g}}_{(\beta,H^{0}_{AB})}-\textbf{\textit{g}}_{(\beta,H_{AB})})],
\end{split}
\end{equation*}
where $S(\boldsymbol{\varrho})=-\mathrm{Tr}[\boldsymbol{\varrho}\ln\boldsymbol{\varrho}]$ and $S(\boldsymbol{\varrho}\Vert \boldsymbol{\sigma})=-\mathrm{Tr}[\boldsymbol{\varrho}\ln \boldsymbol{\sigma}]-S(\boldsymbol{\varrho})$ are, respectively, the von Neumann entropy and the relative entropy. The first inequality can be obtained by expanding $S(\textbf{\textit{g}}_{(\beta,H_{AB})}\Vert \textbf{\textit{g}}_{(\beta,H^{0}_{AB})})$, and the second one is derived by using the fact that $\textbf{\textit{g}}_{(\beta,H_{AB})}$ minimizes the free energy $F_{\beta,H}(\boldsymbol{\varrho})=\mathrm{Tr}[\boldsymbol{\varrho} H] - (1/\beta)S(\boldsymbol{\varrho})$ [or the shorthand notation $F_{\beta}(\boldsymbol{\varrho})$]. Using the triangle inequality and applying the covariance identity (\ref{eq:pertformula}) to the interaction Hamiltonian yields the thermodynamic upper bound on mutual information as a function of local terms of the boundary interaction and their covariances,
\begin{equation}
\label{eq:thermalbound2}
\begin{split}
\mathpzc{I}_{\beta}(A,B)\leqslant &\,\beta^{2}\,\textstyle{\sum_{\mathbbmss{Z} \text{ cross } \partial_{AB}}\sum_{\mathbbmss{W} \text{ cross } \partial_{AB}} \,\int_{0}^{1} ds \int_{0}^{1} dr}\\
&\,\times |\mathrm{cov}^{r}_{\textbf{\textit{g}}_{(\beta,H_{AB}^{s})}}(h_{\mathbbmss{Z}},h_{\mathbbmss{W}})|.
\end{split}
\end{equation}

The $\beta^{2}$ scaling already appears in this equation and does not require the shell summations that follow. It originates from applying the covariance identity (\ref{eq:pertformula}) to the boundary interaction $H_{\partial}$ itself, rather than bounding the trace directly by operator norms. In the standard derivation \cite{wolf2008area}, the trace is bounded by $\Vert H_{\partial}\Vert \Vert \textbf{\textit{g}}_{(\beta,H_{AB}^{0})} - \textbf{\textit{g}}_{(\beta,H_{AB})}\Vert_{1} \leqslant 2\Vert H_{\partial}\Vert$, which discards the temperature dependence of the correlation operator. The covariance identity preserves it, at the cost of a double sum over boundary terms. The shell summation that follows plays a distinct role: it reduces this double sum from $|\partial_{AB}|^{2}$ to $|\partial_{AB}|$, turning what would otherwise be an area-squared law into an area law. Both ingredients are essential and they address different aspects of the bound.
\\

\begin{figure}[tp]
\includegraphics[width=\linewidth]{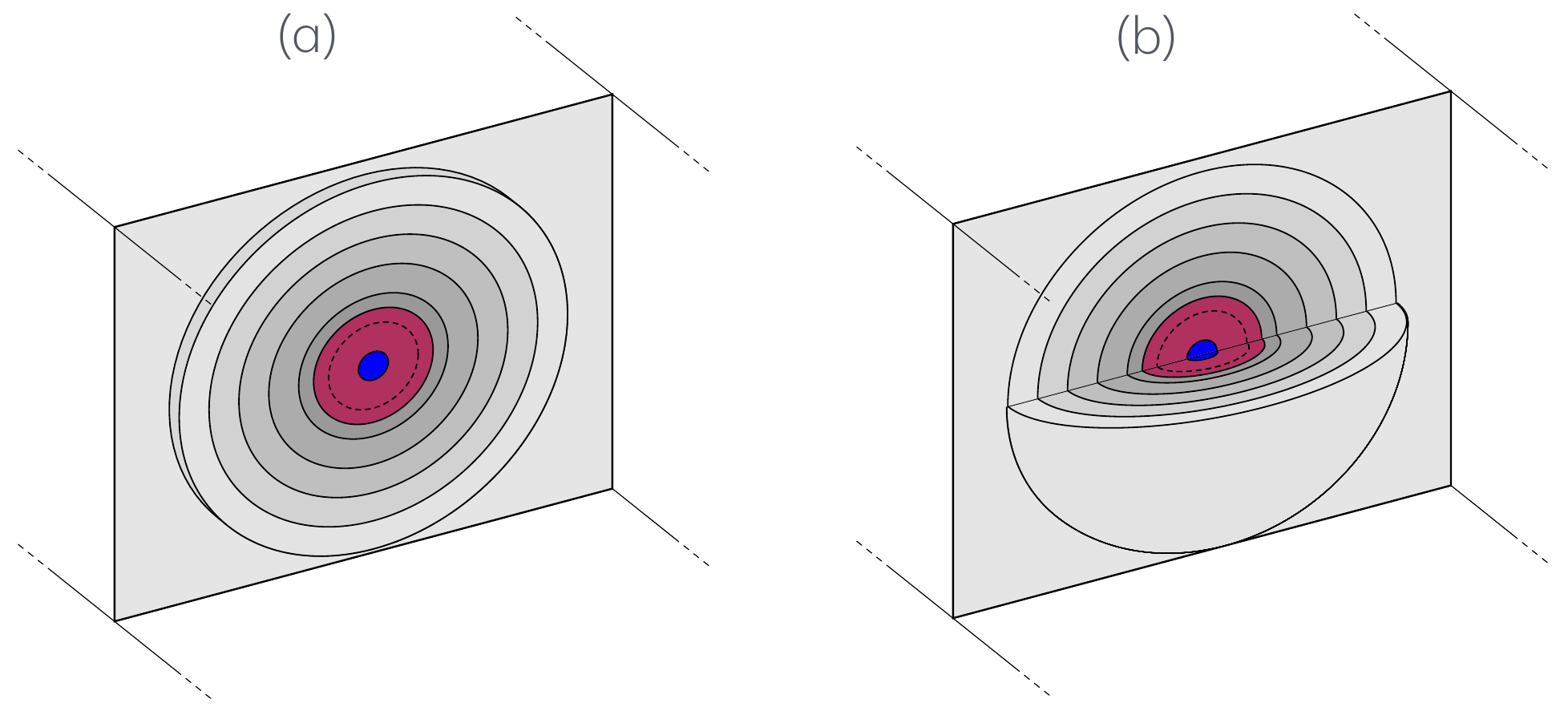}
\caption{Schematic of the $A|B$ boundary and spatial shell summation. (a) The blue and the red disks denotes $\mathbbmss Z$ and the inner core, and the black circles show the $R$-width radial grouping of the local boundary terms $\mathbbmss W\text{ crosses }\partial_{AB}$ with respect to their distance from $\mathbbmss Z$. (b) The blue and the red spheres show $\mathbbmss Z$, and $V_{\mathrm{core}}$, and the black spheres show the $R$-width spatial shell grouping of the local bulk terms $\mathbbmss W\subset\Lambda$.}
\label{fig:shells}
\end{figure}

\textit{Exponential clustering and shell summation.---}Now we employ a seminal observation of Ref. \cite{kliesch2014locality} about clustering of correlations. For local Hamiltonians, one can discern a universal critical inverse temperature $\beta^{*}=\ln\big[(1+\sqrt{1+4/\alpha})/2\big]/2 h$ (related to the convergence radius of the cluster expansion, with $\alpha$ a lattice constant) such that, for all $\beta < \beta^{*}$, the covariance between strictly local observables decays exponentially,
\begin{equation}
\label{eq:thm2}
\hskip-3mm|\mathrm{cov}_{\textbf{\textit{g}}_{(\beta,H^{s}_{AB})}}^{\,r}(h_{\mathbbmss{Z}}, h_{\mathbbmss{W}})| \leqslant C(\beta) \,a_{\mathbbmss{Z} \mathbbmss{W}}\, \Vert h_{\mathbbmss{Z}} \Vert \, \Vert h_{\mathbbmss{W}} \Vert \,e^{-\frac{\mathrm{dist}(\mathbbmss{Z}, \mathbbmss{W})}{\xi(\beta)}},
\end{equation}
provided that $\mathrm{dist}(\mathbbmss{Z},\mathbbmss{W}) \geqslant L_{0}(\beta,a_{\mathbbmss{Z}\mathbbmss{W}})$. Here $L_{0}(\beta,a_{\mathbbmss{Z}\mathbbmss{W}})=\xi(\beta)|\ln[\ln(3)(1-e^{-1/\xi(\beta)})/|\tilde\partial h_{\mathbbmss{Z}}|]|$ is an exclusion radius, $\xi(\beta) =1/|\ln\big(\alpha \, e^{2\beta h} (e^{2\beta h } - 1) \big)|$ is the thermal correlation length, $C(\beta) = 4/[\ln(3)(1 - e^{-1/\xi(\beta)})]$ is a temperature-dependent prefactor, the parameter $\alpha\leqslant [(2R+1)^{D}-1]e$ is a lattice growth constant for lattices of the $\mathds{Z}^{D}$ geometry \cite{kliesch2014locality, Miranda2011latticetrees}, $\tilde \partial h_{\mathbbmss{Z}}$ is the set of all interaction hyperedges overlapping the boundary of $\mathbbmss{Z}$, and we have set $a_{\mathbbmss{Z}\mathbbmss{W}}=\min\{|\tilde\partial h_{\mathbbmss{Z}}|,|\tilde\partial h_{\mathbbmss{W}} |\}$. Since the support of each local term $h_{\mathbbmss{Z}}$ contains at most $R^{D}$ sites and each site is involved in at most $\zeta$ interactions, we can bound this hyperedge boundary by $a_{\mathbbmss{Z}\mathbbmss{W}}\leqslant \zeta R^{D}=:a_{\max}$. 

To evaluate the boundary sum over $\mathbbmss{W}$ in Eq. (\ref{eq:thermalbound2}), we divide the interaction terms $h_{\mathbbmss{W}}$ into an inner core schematically illustrated as the red circle in Fig. \ref{fig:shells} (a) and an outer tail grouped in gray disks spanning all over the rest of the boundary. The exponential clustering (\ref{eq:thm2}) is conditioned to an exclusion radius $L_{0}(\beta,a_{\max})$ within which the exponential bound does not hold. We define the inner core as all terms $\mathbbmss{W}$ within the distance $\mathrm{dist}(\mathbbmss{Z}, \mathbbmss{W})\leqslant L_{0}(\beta,a_{\max})$. For these terms we employ the trivial operator-norm bound for the covariance. Since each site hosts at most $\zeta$ interactions, and the number of the sites in the inner core is bounded by volume of a $D$-dimensional cube of length $R+2L_0$, the number of such terms is bounded as $V_{\mathrm{core}}(\beta)\leqslant \zeta\big(R+2L_{0}(\beta,a_{\max})\big)^{D}$. Note that we are bounding the number of Hamiltonian terms with at least a site in the inner core of the boundary [the red circle in Fig. \ref{fig:shells} (a)] by overcounting over a volume stretched into the bulk [the red sphere in Fig. \ref{fig:shells} (b)]. This adds to the looseness of our bound, however, the exponential decay of correlations is strong enough to suppress any polynomial growth and retains the $\beta^{2}$ behavior. 

For the remaining terms in the outer tail, $\mathrm{dist}(\mathbbmss{Z},\mathbbmss{W})>L_{0}(\beta,a_{\max})$, the exponential clustering applies. We group these outer terms into concentric spatial disks and bound the number of sites on these disks by the volume shells $R_{n,\mathbbmss{Z}}$ around the fixed boundary term $\mathbbmss{Z}$ (compare Figs. \ref{fig:shells} (a) and \ref{fig:shells} (b)), where the normalized distance falls in the $n$th shell $R_{n,\mathbbmss{Z}}=\{h_{\mathbbmss{W}}\,|\,\mathrm{dist}(\mathbbmss{Z},\mathbbmss{W})/R\in [n,n+1), n\in\mathrm{I\!N}\cup\{0\}\}$. We also overcount the terms $n\in\{0,1,\dots,\lfloor L_{0}(\beta,a_{\mathbbmss{ZW}})/R\rfloor\}$ for the sake of analyticity of the bound, hence
\begin{align}
\label{eq:covbound}
&\textstyle{\sum_{\mathbbmss{W} \text{ crosses } \partial_{AB}}}|\mathrm{cov}^{r}_{\textbf{\textit{g}}_{(\beta,H_{AB}^{s})}}(h_{\mathbbmss{Z}},h_{\mathbbmss{W}})| \leqslant h^{2}\, V_{\mathrm{core}}(\beta) +h^{2}\, C(\beta) \nonumber\\
&\,\times a_{\max}\,\big(|R_{0,\mathbbmss{Z}}|+ \textstyle{\sum_{n=1}^{\infty}} |R_{n,\mathbbmss{Z}}|e^{-nR/\xi(\beta)}\big).
\end{align}
The number of terms in the $n$th shell is bounded by the surface area of a cube of length $2(n+1)R+R$ multiplied by width $R$, i.e., $|R_{n,\mathbbmss{Z}}|\leqslant 2\zeta D\big(2(n+1)R+R\big)^{D-1}R$. The idea of bounding shell volumes polynomially in $n$ is inspired by Ref. \cite{barthel2012quasilocality}. Note that for any $n\geqslant 1$ we can substitute $2n+3\leqslant 5n$ to obtain $|R_{n,\mathbbmss{Z}}|\leqslant M n^{D-1}$, where $M=2\zeta D5^{D-1}R^{D}$. For the zeroth shell it is bounded by the volume of a cube of length $3R$, whence $|R_{0,\mathbbmss{Z}}|\leqslant \zeta (3R)^{D}=:R_{0}$. Next, we employ the inequality $y^{a} \leqslant (a / e \delta)^{a} e^{\delta y}$ (valid for all positive $y$, $a$, and $\delta$) and let $y=n+1$, $a=D-1$, and $\delta = R/2\xi(\beta)$, hence 
\begin{equation*}
\textstyle{(n+1)^{D-1}} \leqslant \big( \frac{2(D-1) \xi(\beta)}{e R} \big)^{D-1} e^{(n+1) R / 2\xi(\beta)}.
\end{equation*}
We absorb all $\beta$-dependence into a single analytic prefactor
\begin{equation*}
\begin{split}
\hskip-2mm f(\beta)= \textstyle V_{\mathrm{core}}(\beta) + C(\beta)\, a_{\max}\big[R_0 + \frac{M\left(2(D-1)\,\xi(\beta)/eR\right)^{D-1}}{1-e^{-R/2\xi(\beta)}}\big]. 
\end{split}
\end{equation*}
Substituting this in Eq. (\ref{eq:covbound}) and then back in Eq. (\ref{eq:thermalbound2}) gives the area law (\ref{arealaw-}).

\begin{figure}[tp] 
\includegraphics[width=\linewidth]{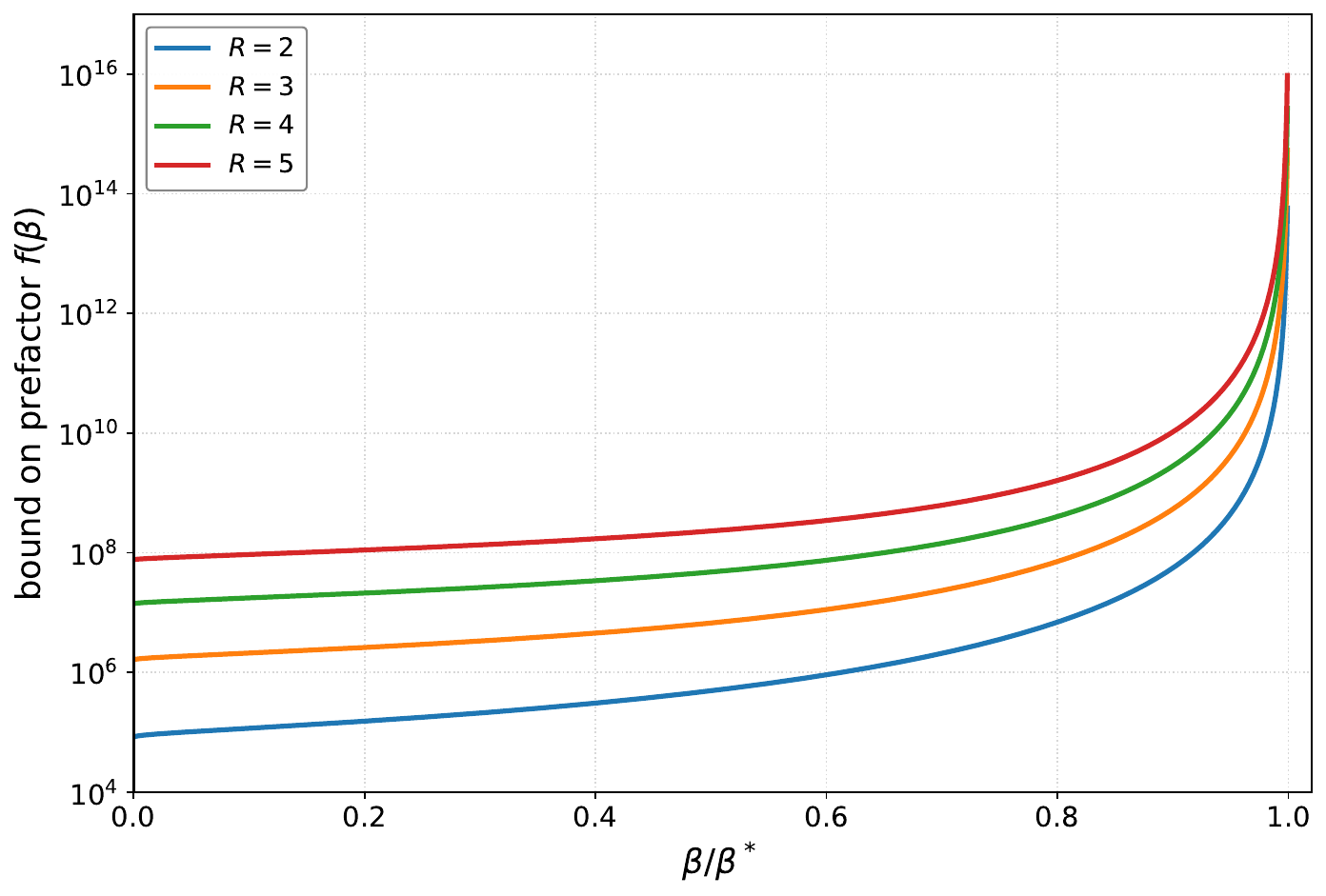}
\caption{A bound on the prefactor $f(\beta)$ vs. relative inverse temperature $\beta/\beta^{*}$ for a $2$-dimensional cubic lattice, for different values of the Hamiltonian range $R$. Note that $V_{\mathrm{core}}$, $|R_{0,\mathbbmss{Z}}|$, and $M$ are replaced by the bounds given in the text, and we have assumed two-body interactions, hence $\zeta=2R(R+1)$.}
\label{fig:prefactor}  
\end{figure}

At the high-temperature limit ($\beta\to  0$), the correlation length and the exclusion radius vanish ($\xi(\beta),L_{0}(\beta,a_{\max})\to 0$) and the infinite spatial sum in Eq. (\ref{eq:covbound}) exponentially collapses to zero. The prefactor stabilizes to a finite geometric constant, $\lim_{\beta\to 0}f(\beta)\leqslant \zeta R^{D}[1+(4/\ln 3)R_{0}]$, confirming that the $\beta^{2}$ scaling is isolated without hidden divergences. In addition, as depicted in Fig. \ref{fig:prefactor}, we observe that as temperature approaches the validity limit $\beta\to \beta^{*}$ the prefactor diverges which diminishes the utility of the bound for that temperature regime, while the traditional linear area law (\ref{lin-AL}) remains valid at all temperatures.
\\

\textit{Correlation bound.---}The Pinsker inequality \cite{Hayashi2016} can be used to bound the distance between the joint thermal state and its marginal factorization, $\Vert \textbf{\textit{g}}_{(\beta,H_{AB})} - \boldsymbol{\varrho}_{A}\otimes \boldsymbol{\varrho}_{B} \Vert_{1} \leqslant \textstyle\sqrt{2\mathpzc{I}_{\beta}(A,B)}$. Now by substituting the area law (\ref{arealaw-}) and using the subsystem thermality (\ref{rhog}) we obtain 
\begin{align*}
&\Vert \textbf{\textit{g}}_{(\beta,H_{AB})} - \boldsymbol{\varrho}_{A} \otimes \boldsymbol{\varrho}_{B}\Vert_{1} \leqslant h\beta\,|\partial_{AB}|\sqrt{2\zeta f(\beta)}, 
\\
&\Vert\textbf{\textit{g}}_{(\beta,H_{AB})} - \textbf{\textit{g}}_{(\beta,H_{A})} \otimes \textbf{\textit{g}}_{(\beta,H_{B})} \Vert_{1} 
\leqslant h\beta\,|\partial_{AB}|\big(\sqrt{2\zeta f(\beta)} + 2\zeta\big). 
\end{align*}
Both bounds scale as $O(\beta\, |\partial_{AB}|)$ at high temperatures and vanish when $\beta \to 0$, showing that the global thermal state asymptotically factorizes.
\\

\textit{Decoupling free energy and extractable work.---}The extractable work from correlations has been studied in the setting of noninteracting, locally thermal subsystems, where it is bounded by the mutual information, $W_{\max} \leqslant \beta^{-1} \mathpzc{I}_{\beta}(A,B)$ \cite{perarnau2015extractable}. Here the correlations originate from a boundary interaction on an interacting lattice, and we bound the work directly through the Gibbs variational principle. To evaluate thermodynamic significance of boundary correlations, consider a dynamical quench where the interaction $H_{\partial}$ is instantaneously turned off, i.e., $H_{AB} \to H^{0}_{AB}$. The composite system, initially in the global Gibbs state $\textbf{\textit{g}}_{(\beta,H_{AB})}$, is now out of equilibrium with respect to $H^{0}_{AB}$. 

If the system is subsequently allowed to thermalize in contact with a bath at inverse temperature $\beta$, the maximum work that can be extracted during this process is bounded by the nonequilibrium free energy difference $W_{\max} = F_{\beta,H^{0}_{AB}}(\textbf{\textit{g}}_{(\beta,H_{AB})}) - F_{\beta,H^{0}_{AB}}(\textbf{\textit{g}}_{(\beta,H^{0}_{AB})})$. 

To bound this work, we recall that the Gibbs state $\textbf{\textit{g}}_{(\beta,H)}$ minimizes the free energy $F_{\beta,H}(\boldsymbol{\varrho})$. Thus, $F_{\beta,H_{AB}}(\textbf{\textit{g}}_{(\beta,H_{AB})}) \leqslant F_{\beta, H^{0}_{AB}}(\textbf{\textit{g}}_{(\beta,H^{0}_{AB})})$ or $F_{\beta,H^{0}_{AB}}(\textbf{\textit{g}}_{(\beta,H_{AB})}) + \mathrm{Tr}[H_{\partial} \,\textbf{\textit{g}}_{(\beta,H_{AB})}] \leqslant F_{H^{0}_{AB}}(\textbf{\textit{g}}_{(\beta,H^{0}_{AB})}) + \mathrm{Tr}[H_{\partial}\, \textbf{\textit{g}}_{(\beta,H^{0}_{AB})}]$. Rearranging this inequality yields a bound on the extractable work as $W_{\max} \leqslant \mathrm{Tr}[H_{\partial}\, (\textbf{\textit{g}}_{(\beta,H^{0}_{AB})} - \textbf{\textit{g}}_{(\beta,H_{AB})})]$. The term on the right-hand side is exactly the quantity we bounded earlier using the covariance and spatial shell summation. Substituting Eq. (\ref{eq:covbound}) summed over the boundary gives Eq. (\ref{wmaxb}). This relation demonstrates that the maximum work extractable from the state after decoupling scales as $O(\beta)$. The thermodynamic value of the correlations is  boundary-limited and decays at high temperatures, restoring compatibility with standard macroscopic thermodynamics.
\\

\textit{Binding energy.---}Consider the deviation of the total internal energy $U_{AB}(\beta)=\mathrm{Tr}[H_{AB}\, \textit{\textbf{g}}_{(\beta,H_{AB})}]$ from the decoupled bulk energy $U^{0}_{AB}(\beta)=\mathrm{Tr}[H^{0}_{AB}\, \textit{\textbf{g}}_{(\beta,H^{0}_{AB})}]$. The binding energy of the interface $\Delta U_{AB} = U_{AB} - U^{0}_{AB}$ consists of the boundary term $\mathrm{Tr}[H_{\partial}\, \textit{\textbf{g}}_{(\beta,H_{AB})}]$ and the correlation term $\mathrm{Tr}[H_{AB}^{0}(\textit{\textbf{g}}_{(\beta,H_{AB})} - \textit{\textbf{g}}_{(\beta,H^{0}_{AB})})]$. Without loss of generality, we assume that the interaction terms $h_{\mathbbmss{Z}}$ are grouped such that they are traceless $\mathrm{Tr}[h_{\mathbbmss{Z}}]=0$. This allows one to rewrite $\mathrm{Tr}[H_{\partial}\,\textit{\textbf{g}}_{(\beta,H_{AB})}]=\mathrm{Tr}[H_{\partial}\, (\textit{\textbf{g}}_{(\beta,H_{AB})} - \textbf{\textit{g}}_{(0,H_{AB})})]$ whence
\begin{align*}
\hskip0mm \text{boundary term} &= \big| \textstyle{\int_{0}^{\beta}} d\beta' \, d\mathrm{Tr}[H_{\partial} \, \textit{\textbf{g}}_{(\beta', H_{AB})}]/d\beta'\big| \nonumber \\
&= \big| \textstyle{\int_{0}^{\beta}} d\beta' \, \mathrm{cov}_{\textit{\textbf{g}}_{(\beta', H_{AB})}} (H_{AB}, H_{\partial}) \big| \nonumber \\
&\leqslant \beta \,\textstyle{\sum_{\mathbbmss{Z} \text{ crosses } \partial_{AB}}} \,\textstyle{\sum_{\mathbbmss{W} \subseteq \Lambda}} | \mathrm{cov}_{\textit{\textbf{g}}_{(\beta,H_{AB})}}(h_{\mathbbmss{W}}, h_{\mathbbmss{Z}})| \nonumber \\
&\leqslant \beta \,\textstyle{\sum_{\mathbbmss{Z} \text{ crosses } \partial_{AB}}} h^{2} f(\beta) \nonumber \\
&\leqslant \zeta h^{2}f(\beta)\, \beta\, |\partial_{AB}|. 
\label{eq:BEbound}
\end{align*}
Note that while the inner sum  ranges over the entire bulk volume $\Lambda$, it is bounded by the exact same analytic function $f(\beta)$. This is because our derivation of $f(\beta)$ in Eq. (\ref{eq:covbound}) utilized the full spatial shell volume [Fig. \ref{fig:shells} (b)], which bounds all interactions at a given distance regardless of whether they lie on the boundary or inside the bulk. Thus, the geometric overcounting natively equips $f(\beta)$ to absorb extensive bulk summations without volume divergences. 

To bound the correlation term (the bulk contribution) in $\Delta U_{AB}$, we set $O=H^{0}_{AB}$ and use the covariance identity (\ref{eq:pertformula}) to calculate $| \mathrm{Tr}[H_{AB}^{0} (\textit{\textbf{g}}_{(\beta,H_{AB})} -\textit{\textbf{g}}_{(\beta,H_{AB}^{0})})]|$. Expanding the Hamiltonians into local terms yields a double summation structurally identical to the boundary energy derivation. We then apply the shell summation bound (\ref{eq:covbound}) which gives the correlation term $\leqslant \zeta h^{2}\,f(\beta)\, \beta\, |\partial_{AB}|$. Hence the binding energy of interface $A|B$ is obtained as in Eq. (\ref{binden}) \cite{note}.
\\

\textit{Summary and outlook.---}We have proved an exact, nonperturbative area law for quantum mutual information of thermal states of local lattice Hamiltonians, which is quadratic in inverse temperature and holds for all inverse temperatures below a clustering threshold temperature, with a prefactor that remains finite as temperature increase and diverges at the threshold value. This is a rigorous improvement over the existing linear bounds and hence provides a sharper control for many-body thermalization. We have also derived corresponding area laws for the maximum work extractable from decoupling of two regions and for the binding energy of the interface, both linear in inverse temperature. These results resolve a thermodynamic inconsistency of the standard treatment of the linear thermal area law, which permits nonvanishing, surface-sized extractable work at arbitrarily high temperatures. The conceptual origin of the improvement is a change of perspective. The common intuition that surface effects do not penetrate far into the bulk would suggest that the temperature dependence of the mutual information is governed by boundary-bulk correlations. Rather, what we have shown is that it is governed by covariances between pairs of boundary interaction terms. The operative intuition is boundary-boundary, i.e., what happens at one point on the boundary does not travel far along the boundary. This perspective complements the low-temperature improvement of the thermal area law, which uses a polynomial approximation to the Gibbs state. 

Since our main result relies only on the locality of the Hamiltonian and the exponential clustering theorem, it is expected that all systems meeting these conditions, such as fermionic lattice systems, feature the same quadratic area law without drastic modifications. The high-temperature bound of this paper and the existing low-temperature result cover two disjoint ranges of temperature. Whether they can be interpolated into a single bound that is optimal (or almost optimal) across the entire temperature range remains an open problem. In addition, the boundary-boundary perspective suggests a natural question for the theory of Gibbs state sampling. The existing low-temperature improvement was accompanied by an improved matrix-product operator approximation, but the derivation proceeds through a polynomial approximation of the Gibbs state rather than through the area law itself. It will be interesting to explore whether our high-temperature improvement can similarly be translated into an algorithmic gain. Our exact results can provide a unified and rigorous foundation bridging quantum information and quantum matter, offering powerful analytic tools for various problems such as efficient simulation and preparation of quantum Gibbs states on quantum computers, quantum thermalization, and probing warm quantum matter and bulk-boundary entanglement geometries in black-hole thermodynamics.
\\

\textit{Acknowledgments.}---This work was supported by Sharif University of Technology's Office of Vice President for Research and Technology through Grant No. QST4040202.


\end{document}